\documentstyle[editedvolume]{crckapb}

\begin{opening}
\title{SUMMARY TALK : MULTI-WAVELENGTH SKY SURVEYS (IAU 179)}

\author{OFER LAHAV}
\institute{Institute of Astronomy, Madingley Road, Cambridge CB3 0HA, UK} 

\end{opening}

\runningtitle{SUMMARY TALK}

\begin{document}


\section{Introduction}

An astronomer's career can be viewed in a 3-dimensional space where 
the (nearly orthogonal) axes are : 
(i) the objects of interest (from planets to the Universe), 
(ii) techniques (from instrument design to analytic calculations),
and  (iii) the wavelength (from the radio to gamma rays).
This interdisciplinary conference brought together experts 
from different bands of the `wavelength axis'. 
It has been an interesting meeting, with a lot of cross-talk,
excellent review talks and high-quality posters.

I shall begin by summarizing the highlights 
(as described in the oral presentations) 
in different wavelengths,
and  then discuss the need for statistical techniques,
some key astrophysical questions which should be addressed 
by multi-wavelength (MW) approaches,
and the changing sociology of the field.

\section {Across the spectrum} 

For most of the history of mankind
our picture  of the universe was 
restricted to the visual band.
However, in particular since the Second World War there 
has been rapid progress
in instrumentation and space technology which allows us  
to have a nearly complete MW picture of astronomical objects
and background radiations.
Below is a brief summary of what we have heard at the conference, 
ordered from long to short wavelengths.

\subsection {RADIO}

Being a `modern' wavelength, many new astronomical phenomena were
discovered in the radio.  In fact, all Nobel prizes in observational
Astronomy were given to radio-astronomers: for the discoveries of the
Cosmic Microwave Background (CMB) Radiation, 
the pulsars, and the binary pulsar.
At this meeting we have mainly heard about radio extragalactic surveys
(Becker, Condon, Disney, Fuerst, Sokolov, Wall).  Present radio
surveys (e.g. 87GB and PMN) have median redshift ${\bar z} \sim 1$,
hence providing an unusually deep picture of the universe at early epochs.
Recent studies have indicated that radio sources are strongly clustered,
suggesting that they reside in high density regions.  Several on-going
and future surveys in the continuum (FIRST, NVSS, WENSS, UTR) are most
promising, although a crucial problem is how to get optical redshifts
to these distant radio objects, in order to map their 3-dimensional
distribution.  The Parkes multibeam survey in 21 cm promises to give
an unbiased view of the neutral hydrogen content in the local universe,
with implications for identification of new galaxy populations
(e.g. dwarfs and low surface brightness galaxies).
Moving to much higher redshifts,
radio measurements of temperature fluctuations on the last scattering 
surface of the CMB by COBE and future experiments 
(COBRAS/SAMBA, MAP, VSA) reveal the seeds of cosmic structure.

\subsection {INFRARED} 

The relatively recent development of detectors in the
infrared provides a new window to 
Galactic and extra-galactic objects 
(as discussed by Beichman, Oliver, Ruphy). 
The IRAS data base (in 12, 25, 60 and 100 micron) 
has  become a major tool for studies of the Galactic structure
and the local universe. In particular, follow-up redshift surveys
(IRAS1.2Jy, QDOT, PSCZ) have yielded  nearly whole-sky 3-dimensional 
catalogues for studies of the density and peculiar velocity fields.
The new  surveys in 2 micron (2MASS, DENIS)
are in particular useful for tracing the old `stable' stellar
population in galaxies, and for overcoming the problem of Galactic extinction.
Together with ISO, they will give a new picture of stellar populations 
within the Galaxy  and a deeper view of large 
scale structure.

\bigskip

\subsection  {OPTICAL} 

This traditional band remains most useful.
Photographic surveys (Djorgovski, Morrison, Reid) 
such as  POSS I+II and  UK Schmidt
still provide the most important wide-angle data bases, 
e.g. as target lists for redshift surveys.
In the old days plates were examined by eye
(e.g. Zwicky, Abell, Nilson, Lauberts), 
but at present the  scanning is done by 
machines (APM, COSMOS, APS, DDS, DPOSS, PMM), with  clever software  
and heroic efforts to calibrate and match plates. 
It is worth stressing that there 
is still no whole uniform sky optical catalogue 
(the nearest to that is a compilation  of UGC, ESO, ESGC, 
followed up  by redshift surveys, e.g. SSRS and  
ORS).

Existing redshift surveys 
(e.g. CfA, SSRS, ORS, IRAS, APM, LCRS, CFRS) 
contain each no more than 30,000
galaxies.  A major step forward using multifibre technology will allow
us in the near future to produce redshift surveys of millions of galaxies.
In particular, two major surveys will probe a median redshift of 
${\bar z} \sim 0.1$.  The
American-Japanese Sloan Digital Sky Survey (SDSS) will yield images in
5 colours for 50 million galaxies, and redshifts for about 1 million
galaxies over a quarter of the sky (see reviews by Bahcall, McKay, 
Szalay).  It will be carried out using a dedicated 2.5m telescope in
New Mexico.  A
complementary Anglo-Australian survey (discussed by Parker/Taylor),
called the 2 degree Field (2dF), will produce redshifts for 250,000
galaxies 
selected from the APM catalogue.  
The survey will utilize a new
400-fibre system on the 4m AAT.
These surveys will probe scales larger than $\sim
30 h ^{-1}$ Mpc.  It will also allow better  determination of
$\Omega$ and bias parameter from redshift distortion.  Surveys like
2dF and SDSS will produce unusually large numbers of galaxy spectra,
providing an important probe of the intrinsic galaxy properties, for
studying e.g. the density-morphology relation.  Several groups
recently devised techniques for automated spectral classification of
galaxies utilising  e.g. Principal Component Analysis
and Artificial Neural Networks.  The SDSS and 2dF projects 
will also carry out important surveys of quasars 
(see report by Smith).

We have heard about progress in 
Astrometry 
(Egret, Jones, Mignard, Seidelman). 
Present and future missions 
(Hipparcos, Tycho, SIM, GAIA, VSOP) 
have various applications to a diverse range of topics: 
e.g. planets, the distance scale and tests of General Relativity.
Gravitational lensing (Cook, Schneider) is another innovative area
where a major part of the activity is in the optical band.
Finally, the impressive  pictures of the Hubble Deep Field 
(reviewed by Dickinson) 
reveal morphologically disturbed galaxies, 
probably in the process of formation.

\subsection {UV, EUV} 

Th UltraViolet  and extreme UV bands (reviewed by Brosch)
have been explored by the missions  
TD1, UIT, GLAZAR and  FOCA
and future projects include e.g.
TAUVAX, MSX and  ARGOS.
Here we only point out two important implications  of the UV band.
The first is that in order to compare local galaxies 
with the images at high redshift (e.g. from the Hubble Deep Field) 
one has to image the local galaxies in the UV.
The second  issue is the relevance  of  detection of
the UV  background, which plays an important role 
e.g. in the evolution of Lyman-$\alpha$ clouds.
It is also worth emphasizing the lack of a deep full-sky survey in 
the UV.

\subsection {X-RAY}

The study of the X-ray band (discussed by Hasinger, Stewart, Trumper)
is less than 40 years old, but it opened a new window to high energy
phenomena.  Satellites like Uhuru, HEAO1, ROSAT, Ginga and ASCA and
future missions (ABRIXAS, AXAF, XMM, SRG) probe a wide range of
astronomical objects in the X-ray, including comets, binary stars,
AGNs, clusters of galaxies and the yet unexplained X-ray Background.
We have learned that 50 \% of the ROSAT resolved objects are AGNs,
10\% are galaxies and clusters, 35 \% stars and 5 \% other objects.
X-ray clusters are important cosmological probes:
the X-ray temperature tells us the depth of the cluster
potential well, and hence the total mass, while the X-ray emission
indicates a large fraction of baryonic mass, with implications
for the density parameter $\Omega$ (see below).
Optical follow up observations of the 
X-ray selected
clusters are important in order to get redshifts 
(so far obtained for 600 clusters). 

\subsection {$\gamma$-RAYS} 

This is the most `energetic' band
(reviewed by Gehrels and  Miller) which
in particular became active following the Compton GRO observations.
GRO detected Galactic objects, 
Blazars 
(which are most probably due to inverse Compton effect in beamed jets), 
and about 1500 $\gamma$-ray bursts.
The origin of  $\gamma$-ray bursts is still a mystery
and a MW identification programme is essential in order to associate them with 
other astronomical objects at known distances.
One current popular theoretical 
idea is that they are due to merging of 2 neutron stars
at cosmological distances. 
It seems less likely that the $\gamma$-ray bursts 
reside within the Milky Way.
A new innovative project (Milgarno)
utilises  air showers/Cerenkov radiation 
to detect very high energy $\gamma$-rays.

\section {Statistical techniques} 

We have heard many talks about cross identification of objects observed
at different wavelengths (Bartlett, Becker, Brunner, Hasinger, Helou,
Prandoni, Szalay, Wagner, Wall, White).  Typical problems are the
accuracy of coordinates and overlap of objects.  
In principle, 
one should take into
account not only the angular proximity, but also the radial distribution
of the objects in the different catalogues.  For example, the median
redshift of the radio survey 87GB is ${\bar z} \sim 1$, 
much deeper
than local optical and IRAS surveys (${\bar z} \sim 0.02$), hence 
it is  not surprising that Condon finds
only 1 \% of sources in common.  One can formulate the 
cross-identification by taking into account this prior probability for
the radial distribution, but it depends to what extent one is prepared
to be `a Bayesian'.
It is also possible to calculate the non-zero lag cross-correlation between 
catalogues.  As most objects are known to be clustered, this could yield 
valuable extra information. An example is cross-correlation of the 
diffuse X-ray Background with known galaxies. 

Other statistical issues discussed at the meeting are 
selection effects in catalogues (Disney), 
and the need for innovative  methods of classification and 
pattern recognition (Feigelson, Harwit).
Clearly we should  keep an open eye  on developments of 
statistical tools in other fields, 
as many similar problems have already been solved.
Special attention was given at this meeting to 
data bases of astronomical objects and user-friendly software packages 
(e.g. ADS, CDS, SIMBAD, NED, ALADIN, Skyview) 
and on how to handle Terabytes of data (Szalay, White). 
The progress is impressive, and we are all  thankful to those
who compile the data sets and develop public domain software.

\section {Astrophysical Problems }

Here we outline several key problems in Astrophysics which
are best tackles by 
MW approaches (with a bias of the reviewer towards cosmological
problems).  

\subsection{Galactic Structure}

The MW measurements and new techniques such as microlensing 
have renewed interest in the structure and evolution on the Milky Way.
Many speakers 
(Bienayme, Boulanger, Fuerst, Fukui, Gehrels, Parker, Price, Larsen, 
Majewski, Mendez, Ruphy) 
discussed  unsolved problems related to Galactic 
structure and dynamics. Here we only list some of them: 
what is the stellar initial mass function ? 
what is the  extent of the thin and thick discs ? 
is the halo lumpy ?
does the Milky-Way have a bar ? 
how strong is the interaction between the Milky-Way
and nearby dwarfs ?

\subsection {Galaxies behind the Zone of Avoidance}

Recent discoveries of galaxies (e.g. the Sagittarius dwarf and Dwingeloo 1)
and clusters (e.g. A3627 at the centre of the Great Attractor) illustrated 
how a combination of eye-balling of plates, 21 cm, infrared and X-ray 
measurements can be combined to unveil galaxies hidden behind the 
`Galactic fog'.
Although this topic was not directly discussed at this meeting, it is clear
that  collaboration of Galactic and extragalactic astronomers on this topic
can benefit both groups. There is also  a great need
for a better extinction map of the Galaxy 
(e.g. by combining optical, UV, HI, IRAS and COBE 
maps). We were reminded that there is substructure in the ISM even at the 
north and south Galactic poles.

\subsection {Galaxy Formation}

A MW approach is required to produce the `H-R diagram' of galaxies.
Rather than talking just about a luminosity function in one colour,
one should consider a multivariate function which includes
luminosities in various spectral bands and dynamical properties of
galaxies.  Deeper galaxy images such as the HDF hold the key to issues of
galaxy formation and evolution (e.g. Dickinson's review), but to make
sense of the high redshift measurements it is crucial to improve our
knowledge of galaxy properties in the local universe.  Another
constraint on galaxy formation can be obtained 
from the `maximum' redshift of QSOs (Osmer,
Padovani). 

\subsection {Large Scale Structure \& Cosmology}

Recent work on 
understanding 
clusters of galaxies is a good example of  
MW measurements. Optical, X-ray, CMB (for the Sunyaev-Zeldovich 
effect) and gravitational lensing are combined to quantify the extent 
and  matter content of clusters  
(discussed by Bahcall, Trumper, Schneider).
For example,
the high fraction 
of baryons in clusters suggests that 
$\Omega \approx 0.2$, in conflict with  
the popular $\Omega=1$ value.
The derivation of $\Omega$ from comparison of density and velocity fields
and from redshift distortion is affected by 'biasing', i.e. the way 
galaxies trace the underlying mass distribution.
Galaxies observed at different wavelengths (e.g. optical and IRAS)
have different clustering properties, hence a different `bias parameter'.
This conceptual issue will remain  crucial in the interpretation 
of the 2dF and SDSS surveys, and in connecting the 
power-spectrum of density fluctuations from galaxy surveys and the CMB.

MW  studies are also important for understanding the 
background radiations. 
Most of the background energy is in the CMB (0.25 $eV/cm^3$),
compared with  e.g.  the XRB ($5 \times 10^{-5}  eV/cm^3$).
Only upper limits (from direct measurements) 
are available for the optical and UV backgrounds.
An interesting  example of MW approach is how the hot intergalactic medium 
model for the XRB 
was ruled out by the lack of distortion of the black-body CMB spectrum.
It is also important to detect the dipole in various background radiations,
to confirm that the  CMB dipole is due to motion.
Sub-degree measurements of fluctuations in the CMB by 
new experiments (COBRAS/SAMBA, MAP, VSA) will provide a new insight 
into the nature of the dark matter, and will yield 
estimation (in a model-dependent way) of the cosmological parameters 
(e.g. $\Omega$  and $H_0$) to within a few percent.

\section {Changing Sociology}

We have heard at this meeting about  future big telescopes, big
collaborations and big data bases.  Several  speakers
made the remarks  ``it takes at least 2-3 times the most pessimistic estimate to
begin/complete/analyse a survey" and ``to make a big impact, a new
survey must be 10-1000 times better in sensitivity/resolution/number
of objects". This  indicates  that it is becoming more and more challenging to
make an impact by conducting big surveys.  It also raises some
questions about the changing sociology of doing research in astronomy:
what will be the individual's contribution in a big collaboration 
(cf. particle physics experiments) ?
what skills should be acquired by the next generation of
astronomers ?
Will the increase in projects and data sets be followed by
more jobs for young astronomers ?   How
would the community deal with public domain data 
(e.g. HDF) ?  and how to communicate the
knowledge resulting from the surveys  to the tax payer ?
While focusing on the technological aspects of the MW
surveys, the human aspects should not be forgotten.
\bigskip
\bigskip

Finally, many thanks to the scientific and local organizing committees, 
in particular  
to Barry Lasker and Marc Postman, for organizing this  interesting and 
stimulating meeting.

\end {document}